\documentclass{raa_twocolumn}
\usepackage{graphicx,times}             
\usepackage{natbib}
\usepackage{amssymb,amsmath}
\usepackage{multirow}
\usepackage{float}
\usepackage{siunitx}
\bibpunct{(}{)}{;}{a}{}{,}
\usepackage[pagebackref=true]{hyperref}

\begin{document}

  \title{Study on the detector energy response of SVOM/GRM
}

   \volnopage{Vol.0 (202x) No.0, 000--000}      
   \setcounter{page}{1}          

   \author{Xiao-Yun Zhao 
      \inst{1,*}\footnotetext{$*$Corresponding Authors, these authors contributed equally to this work.}
   \and Jiang He  \inst{1}
   \and Shi-Jie Zheng   \inst{1,2,*}
   \and Ping Wang \inst{1,2}
   \and Shao-Lin Xiong \inst{1,2,*}
   \and Yue Huang \inst{1,2}
   \and Dong-Ya Guo \inst{1,2}
   \and Juan Zhang \inst{1,2}
   \and Rui Qiao \inst{1,2}
   \and Hao-Li Shi \inst{1}
   \and Lu Li \inst{1,2}
   \and Li Zhang  \inst{1}
   \and Jin Wang  \inst{1}
   \and Meng Bai \inst{3} 
   \and Yong-Wei Dong \inst{1,2}
   \and Min Gao \inst{1}
   \and Louvin Henri \inst{4} 
   \and Ulysse Jacob \inst{5} 
   \and Yong-Ye Li \inst{1,6}
   \and Jiang-Tao Liu \inst{1,2}
   \and Xin Liu \inst{1,2}
   \and Qing-Yun Mao \inst{7}
   \and Frédéric Piron \inst{5}
   \and Li-Ming Song \inst{1,2}
   \and Rui-Feng Su \inst{7}
   \and Jian-Chao Sun \inst{1,2}
   \and Wen-Jun Tan \inst{1,2}
   \and You-Li Tuo \inst{8} 
   \and Chen-Wei Wang \inst{1,2}
   \and Jin-Zhou Wang \inst{1}
   \and Rui-Jie Wang \inst{1}
   \and Bo-Bing Wu \inst{1,2}
   \and Wen-Hui Yu \inst{1,9}
   \and Shuang-Nan Zhang \inst{1,2}
   \and Shu-Min Zhao \inst{1,10}
   }

   \institute{State Key Laboratory of Particle Astrophysics, Institute of High Energy Physics, Chinese Academy of Sciences, Beijing 100049, China; {\it xyzhao@ihep.ac.cn,  zhengsj@ihep.ac.cn, xiongsl@ihep.ac.cn } \\
        \and 
             University of Chinese Academy of Sciences, Chinese Academy of Sciences, Beijing 100049, China\\
        \and 
            National Space Science Center, Chinese Academy of Sciences, Beijing 100190, China\\
        \and  
        Commissariat à l'énergie atomique et aux énergies alternatives (CEA), Gif-sur-Yvette, France\\
        \and 
        Laboratoire Univers et Particules de Montpellier, Université Montpellier, CNRS/IN2P3, F-34095 Montpellier, France
        \and 
         School of Information Engineering, Nanchang University, Nanchang 330031, China.\\
        \and 
            Innovation Academy for Microsatellites of Chinese Academy of Sciences, Shanghai 201203, China\\
        \and 
        Institut für Astronomie und Astrophysik (IAAT), Eberhard Karls Universität Tübingen, Tübingen, Germany\\
        \and 
        Xiangtan University, Xiangtan 411105, China\\
        \and 
         Tangshan Institute, Southwest Jiaotong University, Tangshan 063000, China  \\
\vs\no
   {\small Received 202x month day; accepted 202x month day}}

\abstract{ The SVOM mission is specifically designed to for the detection and localization of Gamma-Ray Bursts (GRBs) and subsequent follow-up observations. Among the four telescopes installed on the SVOM satellite, the Gamma-Ray Monitor (GRM) plays a crucial role in capturing the prompt emission of GRBs due to its wide field of view (FOV) and broad energy range. Accurate determination of the detector's energy response is vital for analyzing GRM data, particularly considering the significant impact of the atmospheric albedo effect on this response.
This research focuses on deriving the detector's energy response and establishing a calibration database for the GRM, with particular emphasis on investigating the atmospheric albedo effect. The study shows that the contribution of albedo photons to the detector’s effective area depends strongly on the orientation of the GRD line of sight (LoS) relative to Earth and on the incident direction of the GRB.
When the GRD LoS is anti-Earth oriented, the albedo effect is minimal,  with the highest proportion of albedo effective area accounting for approximately 10\% of the total effective area. This occurs when the incident angle of the GRB is nearly perpendicular to the LoS.
Conversely, if the GRD LoS is not pointing away from Earth and the GRB arrives from angles greater than about 90°, the albedo component can become predominant, contributing up to around 100\% of the total effective area. This is especially pronounced in the 8 - 20 keV range, where the direct effective area drops to zero due to the large GRB injection angle. Our results show that, it is necessary for GRM to consider the atmospheric albedo effects in detector response, otherwise the spectral and localization analyses will result in biased measurements.
}

   \authorrunning{X.-Y. Zhao, J. He \& S.-J. Zheng}            
   \titlerunning{Study on the detector energy response of SVOM/GRM}  

   \maketitle
%
%
\section{Introduction}           
\label{sect:intro}

The Space-based multi-band astronomical Variable Objects Monitor \citep[SVOM,][]{wei2016deep} is a collaborative mission between China and France, led by the Chinese National Space Agency (CNSA) and the Centre National d’Etudes Spatiales of France (CNES). It is dedicated to the detection and investigation of gamma-ray bursts (GRBs). SVOM was launched on June 22, 2024, into an orbit at an altitude of 625 kilometers with an inclination of 30 degrees (\citealt{10.1117/12.2311710},\citealt{2020ChA&A..44..269Y}). Designed for a nominal operational lifetime of three years, SVOM may be potentially extendable for an additional two years. 

\begin{figure}[H]
    \centering
    \includegraphics[width=1.0\linewidth]{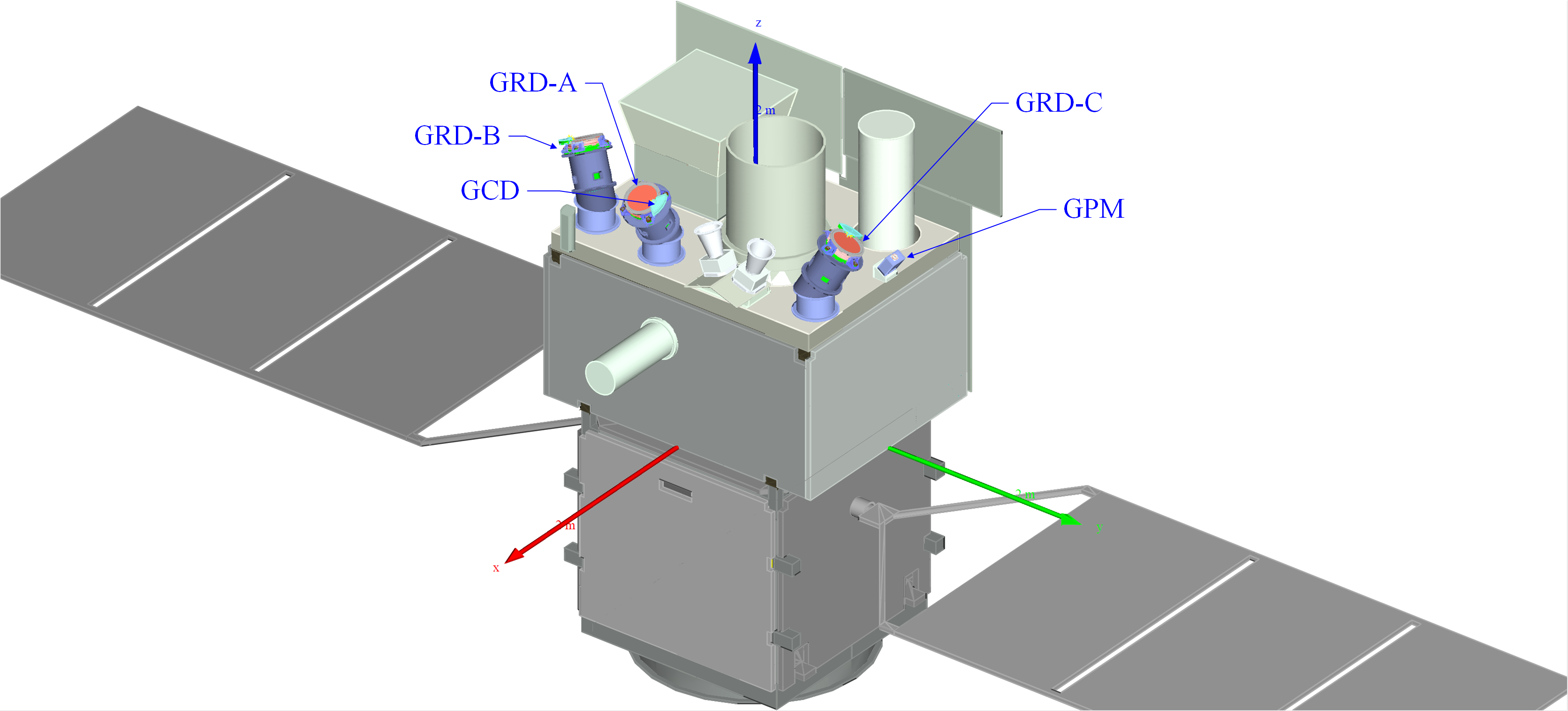}
    \caption{The GRM and its Line of Sight coordinate frame.  }
    \label{Fig1}
\end{figure}

The SVOM mission is equipped with a range of multi-wavelength payloads, comprising two wide-field instruments designed to detect Gamma-Ray Bursts (GRBs) and analyze the spectral characteristics of the prompt emission, known as the GRM, spanning from hard x-ray to soft gamma-ray wavelengths (\citealt{10.1117/12.2313821}, \citealt{2010SCPMA..53S..40D}). Additionally, the mission features the ECLAIRs instrument operating in the hard x-ray band (\citealt{10.1117/12.2055507}) and two narrow-field instruments, namely the MXT for X-ray observations and the VT for optical observations (\citealt{10.1117/12.925171}, \citealt{10.1117/12.2630249}, \citealt{10.1117/12.2561854}). Apart from these onboard instruments, the SVOM mission also integrates various ground-based telescopes, including two 1-meter class robotic telescopes named C-GFT and F-GFT/COLIBRI, as well as a network of wide-field visible cameras known as GWACs. Specifically, C-GFT is situated at the Jilin Observatory in China, COLIBRI is located in San Pedro M\'artir, Baja California, Mexico, and GWACs are positioned at the Xinglong Observatory of NAOC in China.

The recent detection of the extraordinarily bright gamma-ray burst GRB 221009A has underscored the critical importance of precise instrument response calibration for spectral and temporal analysis. Observations from missions like Insight-HXMT and GECAM  have revealed its exceptional brightness and complex spectral features across a broad energy range (\citealt{Zhang2025InsightHXMTOO}, \citealt{2024SCPMA..6789511Z}, \citealt{2024ApJ...962L...2Z}, \citealt{Zhang_2024}). Given that SVOM/GRM operates in the keV–MeV energy range, accurate calibration of its instrument response is essential for reliable physical interpretation.

\begin{figure}[H]
    \centering
    \includegraphics[width=0.6\linewidth]{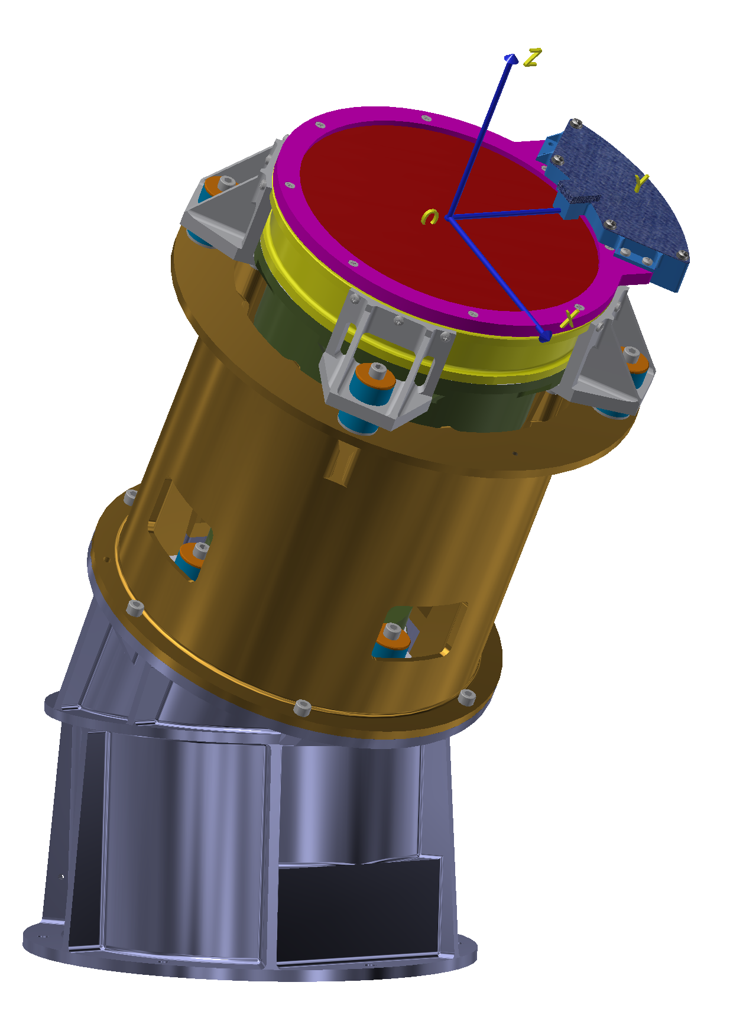}
    \caption{One GRD probe and its Line of Sight coordinate frame.}
    \label{Fig2}
\end{figure}

As one of the two wide-field instruments on board the SVOM satellite, Gamma-Ray Monitor (GRM, Fig.~\ref{Fig1}) is designed to detect gamma-ray bursts in the energy range of 15 keV to 5 MeV (Jian-Chao Sun et al., RAA, 2025, 25, this issue, \citealt{2021NIMPA100365301W}). The GRM comprises three identical Gamma-Ray Detectors (GRDs), denoted as GRD-A/GRD01, GRD-B/GRD02, GRD-C/GRD03 (see Fig.~\ref{Fig2}), mounted on the top panel of the platform. Each GRD is oriented at an elevation angle of 30$^{\circ}$ relative to the LoS axis of the SVOM, with a separation of 120$^{\circ}$ azimuthally between each pair of GRDs (\citealt{mission}). Each GRD is composed of a NaI(Tl) crystal with diameter of 160 mm and thickness of 15 mm, along with a photomultiplier tube (PMT) that connects to the crystal. GRM has the onboard trigger and localization function, the trigger etc. information will be transmitted via the VHF (Very High Frequency) system or the Global Short Message Communication (GSMC) service of BeiDou Navigation Satellite System (BDS) (\citealt{LI20211701}) in near real time (Jiang He et al., RAA, 2025, 25, this issue). The main characteristics of the GRM are summarized in Table ~\ref{Tab1}.

GRM is capable of measuring the temporal, spectral and positional properties of GRBs. Establishing the energy response and creating a CALibration DataBase (CALDB) are critically important for GRM measurements. Gamma-rays emitted by GRBs can directly irradiate the GRD, while some may scatter within the satellite structures. This phenomenon is described by the direct Detector Response Matrix (dDRM). In addition, GRB gamma-rays can interact with the the Earth's atmosphere, where they undergo absorption, attenuation, and reflection \citep{2005NCimC..28..797H}. A portion of the reflected gamma-rays is re-detected by the GRDs. contributing to what is known as the albedo Detector Response Matrix (aDRM). 

 \citet{palit_revisiting_2021} showed that reflections from the Earth's atmosphere can mimic a thermal component predicted by the GRB fireball model. This atmospheric albedo effect plays a critical role in influencing the energy response of the detector. Both Fermi/GBM and GECAM have modeled this effect in their response matrices \citep{2005NCimC..28..797H, SSPMA-2020-0015}. Consequently, a comprehensive understanding of the contribution of the reflected gamma rays is essential for the accurate interpretation of GRM observation data.

 In this study, we establish the CALDB of GRM, with emphasis on incorporating atmospheric albedo effect. We also analyze the effective area attributable solely by the atmospheric albedo effect, providing a reference for spectral and localization analyses. 

%
\begin{table}[H]
\begin{center}
\caption[]{ The main character of GRM.}\label{Tab1}


\begin{tabular}{lc}
\hline\noalign{\smallskip}
Character      &  Value                \\
\hline\noalign{\smallskip}
Crystal  & NaI (Tl)                   \\ 
radius  &  160 mm                      \\
thickness  & 15 mm                     \\
Energy Range & 15-5000 keV              \\
\hline

\end{tabular}
\end{center}
\end{table}

\section{CALDB and Detector energy response}
\subsection{Introduction of CALDB}
\label{subsect:caldb}
\begin{figure}[H]
   \centering
   \includegraphics[width=0.5\textwidth, angle=0]{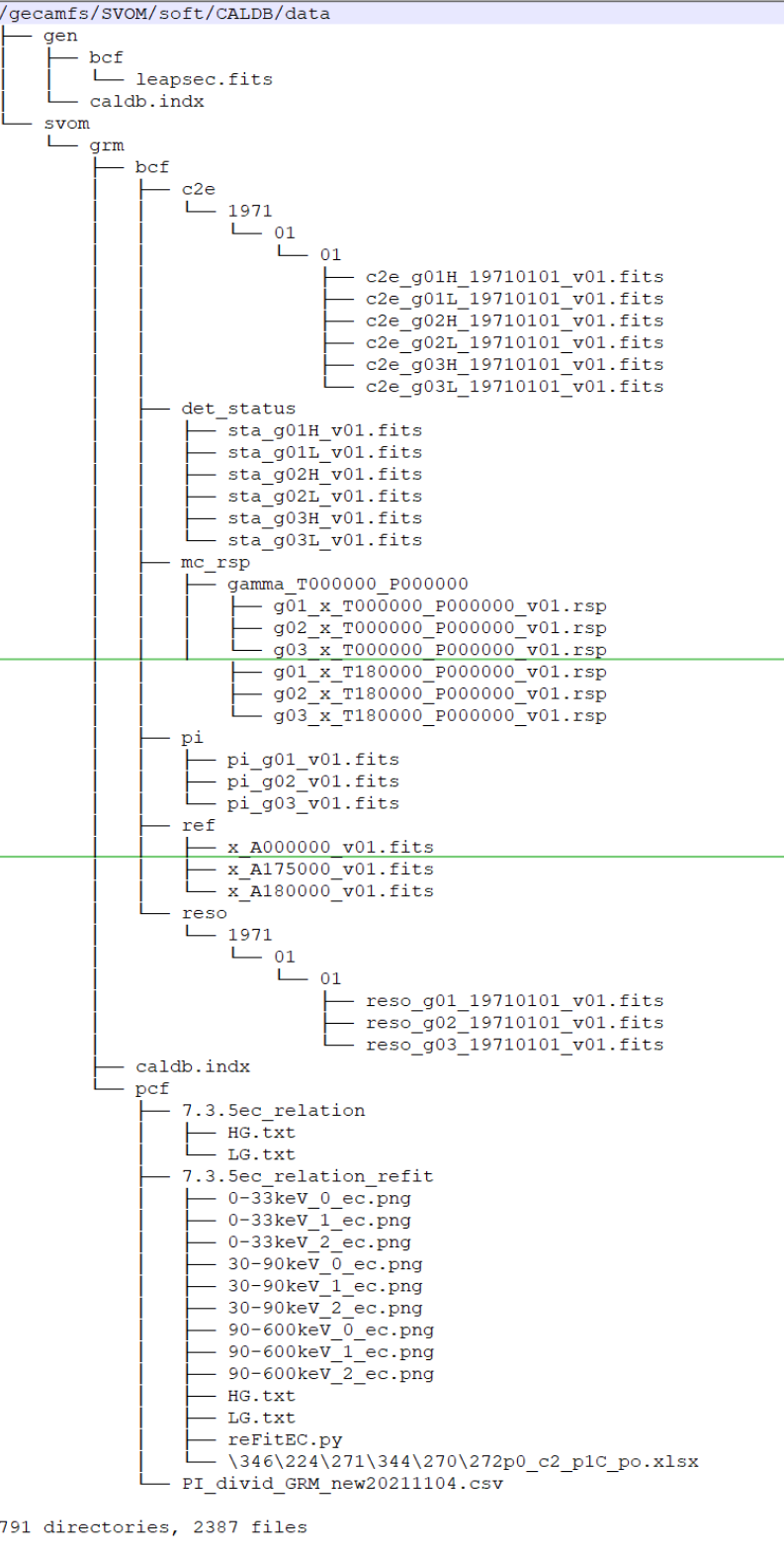}
   \caption{The structure of the GRM CALDB and the file naming rules. (To make it easier to see the entire file structure, the two in green omit most of the similar files in mc\_rsp and ref) }
   \label{Fig2b}
   \end{figure}

Based on the on-ground detailed calibration results and validated simulation results (Jiang He et al., paper in preparation), we have established the CALDB for GRM according to the HEASARC standards. It consists of the calibration data files and the related software package written in python. The total volume of the CALDB is about 80 GB, with a minor portion of about hundreds of MB to be updated regularly in orbit. The structure of the CALDB and the file naming rules are shown in Fig.~\ref{Fig2b}. 

The basic calibration files (bcf) are located in the bcf directory. Within this directory, there are 6 categories and an index file.

The {\bf c2e}  directory contains the Energy and Channel (EC) relationship files. 

The {\bf det\_status} directory stores the lower and higher thresholds for the High gain and Low gain of the GRDs, respectively.

The {\bf mc\_rsp} holds the Monte Carlo simulated detector response files. These files have been validated by on-ground calibration. In the file names, T and P represent the theta and phi angles in the GRM coordinate frame (see Fig.~\ref{Fig1}).

The {\bf pi} directory stores the pulse invariant definition for the GRDs. 

\begin{figure}[H]
\centering
\includegraphics[width=0.4\textwidth]{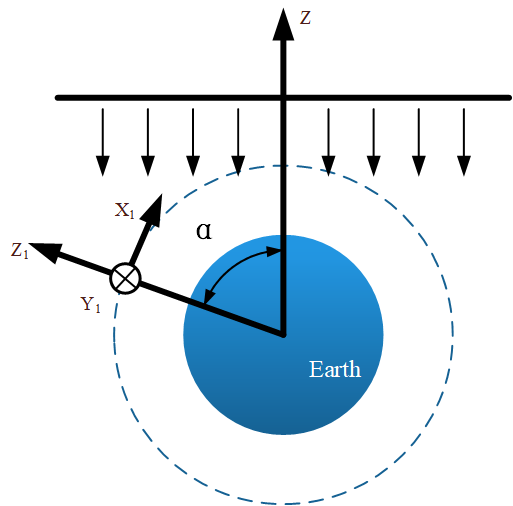}
\caption{The geometry in the atmospheric albedo simulation. $X_{1}Y_{1}Z_{1}$ is the coordinate frame for the scattered photons, where the $+X_{1}$-axis points towards the GRB along the geodesic tangent, the $+Z_{1}$-axis points in the anti-geodesic direction, and the $+Y_{1}$-axis follows the right-hand rule. The origin of the $X_{1}Y_{1}Z_{1}$ coordinate system is the satellite. The black line at the top represents the incident circular surface of the GRB. The angle between the GRB and the satellite, with respect to the center of the Earth, is denoted as $\alpha$.
}\label{Fig3}
\end{figure}

The {\bf ref} directory contains files for the atmospheric albedo factor, organized according to the angle \textbf{A} -- defined as the angle between the SVOM satellite and the source as seen from the center of the Earth (see Fig.~\ref{Fig3}). The files cover the range from 0 to 180° in steps of 5 degrees, resulting in a total of 37 files. Each file is in Flexible Image Transport System (FITS) format with a MATRIX extension and consists of 5400 rows and 3 columns. The columns provide: the theta/phi value of the scattered photon in the scattered photon coordinate frame, and the atmospheric albedo factor derived from simulation (as detailed in subsection \ref{subsect:factor}). The albedo factor is stored as a 300×300 matrix for each angle set, corresponding to logarithmic bins of incident energy and deposited energy, both spanning the range from 5 keV to 40 MeV.

The {\bf reso} directory stores the Energy resolution matrix for each GRD. Using the energy resolution function provided by ground calibration (Jiang He et al., paper in preparation), we calculated the resolution over a deposited energy range of 5 keV to 10 MeV. Then we employed the Gaussian distribution model to generate an energy resolution matrix file that is applicable for both high-gain and low-gain modes. 

A more detailed description of the CALDB can be found here\footnote{https://ihepbox.ihep.ac.cn/ihepbox/index.php/s/je7SapDDNUqPvWp}.

\subsection{Simulation of MC\_RSP}
In the simulation of MC\_RSP (Monte Carlo Response Files), we have constructed the mass model of SVOM in GEANT4. Then, in the GRM coordinate system (see Fig.~\ref{Fig1}), the angles $\theta$ and $\phi$ are divided into 768 groups according to the HEALPix tool. Additionally, three special angles, namely (0°, 0°), (180°, 0°) and (30°, 0°), are added, resulting in a total of 771 angles. Subsequently, gamma photons with a uniform spectrum are incident within the energy range of 5 keV - 40 MeV. The energy range is logarithmically divided into 300 bins. Finally, the deposited energy is collected in 3 GRD detectors. The deposited energy range is 5 keV - 10 MeV, which is divided into 4096 bins using logarithmic spacing. In this way, for each detector and each direction, a DRM file is obtained. The values of Theta and phi are stored in the DRM file name, while the response matrix is stored as one extension of the FITS file.

\subsection{Simulation of Atmospheric Albedo Factor/Effect}
\label{subsect:factor}

In the simulation of the atmospheric albedo factor/effect, a model was developed using GEANT4, which represents the Earth as being surrounded by a 600-km-thick atmosphere composed of 32 concentric shell layers with varying thicknesses. Each layer possesses distinct temperature, pressure, composition, and density parameters, designated according to the NRLMSISE-00 atmospheric model \citep{NRLMSISE}. In GEANT4, the particle source is modeled as a circular surface with a radius of approximately 7000 km. This source emits gamma rays from GRBs in the -Z direction, which corresponds to downward emission toward the Earth (as illustrated in Fig.~\ref{Fig3}). The gamma rays scattered by the atmosphere are recorded by a detector located at an altitude of 600 km. For each scattered gamma ray, its energy and direction are noted. The scattering factor for different scenarios, determined by varying angles of the GRB and the satellite in relation to the Earth's center (represented as $\alpha$), is statistically derived through a uniform methodological approach. For each scenario, corresponding to a specific $\alpha$, a separate file is generated. The simulation covers a range of $\alpha$ from $0^{\circ}$ to $180^{\circ}$, with a step size of $5^{\circ}$, resulting in a total of 37 atmospheric albedo files (refer to one specific file below \textbf{ref} in Fig.~\ref{Fig2b})

\subsection{Generation of an Atmospheric Scattering Matrix}
\label{subsect:M}

In the calculation of a Detector Response Matrix (DRM) with considering the atmospheric scattering effect, the atmospheric scattering matrix is needed. We get this matrix as the following: An interpolation of the atmospheric albedo factor files (these files are retrieved from the \textbf{ref} directory of CALDB, as detailed in subsection \ref{subsect:caldb}) is performed to generate the corresponding factor file for a specific geometrical scenario, which will be utilized in the calculation of the atmospheric scattering matrix $M$. To compute $M$, the existing mc\_rsp is employed. Specifically, for each group of the 5400 angles in this interpolated atmospheric albedo factor file, the corresponding vector is transformed from the scattered photon coordinate frame (refer to Fig.~\ref{Fig3}) into the GRM LoS coordinate frame (as shown in Fig.~\ref{Fig1}) via the J2000 coordinate frame, to find the target mc\_rsp corresponding to the inverse direction of this group of scattered photons. The atmospheric albedo factor matrix is then multiplied by the target mc\_rsp, then the results for all 5400 directions are summed. Considering all the above processes, $M$ can be expressed as follows:
\begin{multline}
M(\alpha,E_{r},E_{m}) = \\
\sum_{i,j}^{5400}F(\alpha(xyz,ra,dec),qra(\theta_{i},\phi_{j}),qdec(\theta_{i},\phi_{j}),\\
E_{r},E_{s}) 
\times mc\_rsp(\theta_{i}^{'},\phi_{j}^{'},E_{s},E_{m}).
\label{Eq1}
\end{multline}

\noindent The details of Equation \ref{Eq1} are listed below:
\begin{enumerate}
\item $\alpha(xyz,ra,dec)$ denotes the angle between the GRB and the satellite with respect to the Earth center (i.e., the angle between $Z_{1}$ and $Z$ in Fig.~\ref{Fig3}), $xyz$ represents the satellite's position in J2000, while $ra$, $dec$ denotes the J2000 coordinates of the GRB.

\item $qra(\theta_{i},\phi_{j})$ and $qdec(\theta_{i},\phi_{j})$ indicate the azimuth and zenith angles of the photon in the scattered photon coordinate frame (refer to Fig.~\ref{Fig3}). $qra$ ranges from [0, $360^{\circ}$) in increments of $2^{\circ}$, while $qdec$ ranges from [0, $180^{\circ}$) in increments of $3^{\circ}$, resulting in a total of 5400 combinations.

\item $\theta_{i}$ and $\phi_{j}$ represent the zenith and azimuth angles, respectively, of the scattered photon in the GRM coordinate frame. 

\item $\theta_{i}^{'}$ and $\phi_{j}^{'}$ represent the zenith and azimuth angles, respectively, for the direction that is antipodal to the direction specified by $\theta_{i}$ and $\phi_{j}$ in the GRM coordinate system.

\item $E_{r}$ is the energy of the incident GRB photon, with a bin number set to 300.

\item $E_{s}$ is the energy of the GRB photon after scattering by the atmosphere, also with a bin number of 300.

\item $E_{m}$ is the energy of the photon as detected by the GRD. 

\item $F$ denotes the scale factor of the scattered photon emerging from a specific direction with a certain energy; this factor is obtained from GEANT4 simulations and stored in the atmospheric albedo factor file. 

\item $i$ and $j$ denote the angular bins of the scattered photon.
\end{enumerate}

\subsection{Generation of a DRM}

\begin{figure}[H]
  \begin{minipage}[t]{0.7\linewidth}
  \centering
   \includegraphics[width=70mm]{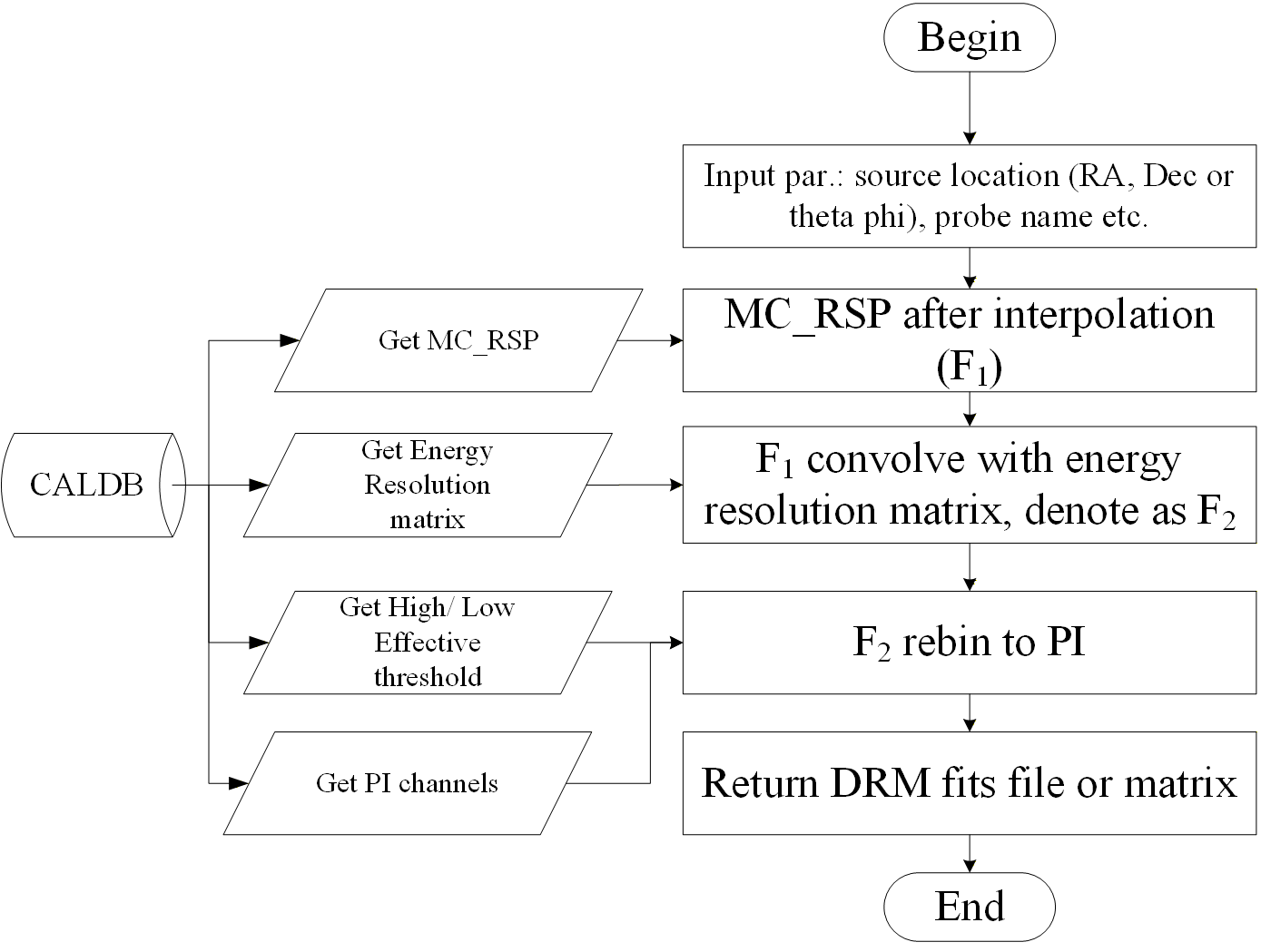}
   \end{minipage}
     \begin{minipage}[t]{0.7\linewidth}
  \centering
     \includegraphics[width=70mm]{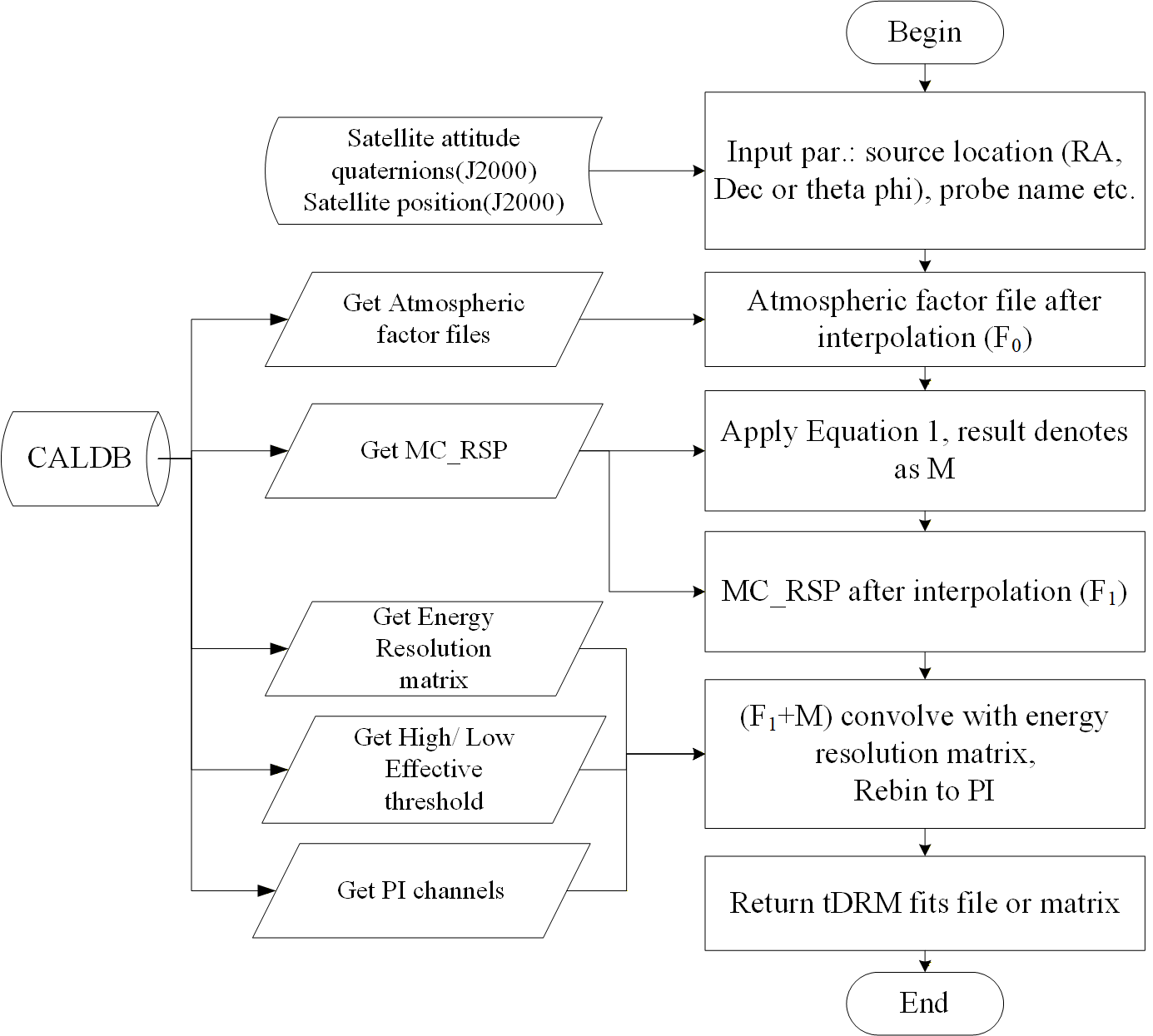}
   \end{minipage}

 \caption{
 The procedure for generating a DRM file, without (top panel) or with (bottom panel) the consideration of atmospheric albedo effects.
 }
 \label{Fig4}
\end{figure}

In the GRM CALDB, a DRM file can be generated either with or without considering the atmospheric albedo effect (Fig.~\ref{Fig4} illustrates the procedure).

When the atmospheric albedo effect is not considered, the required input parameters include the GRB position (specified either as theta/phi or RA/Dec), the probe name (e.g. GRD01), the event type (such as 'evt' for event data or 'vhf' for VHF data), and the GRB trigger time. Based on the GRB's theta/phi, we first perform interpolation using the corresponding MC\_RSP files of the three nearest angles. The resulting interpolated matrix, denoted as $F_1$, is then convolved with the energy resolution matrix to produce $F_2$. Finally, after retrieving of the high and low effective energy thresholds from the CALDB, $F_2$ is rebinned into PI channels. Simultaneously, the deposited energy range is trimmed to align with the effective upper and lower thresholds.

When the atmospheric albedo effect is enabled ('consAtmEff' set to True), calculating $\alpha$ requires additional inputs: the satellite's J2000 position and its attitude quaternions in the J2000 frame. For a given $\alpha$, we perform an interplation using the atmospheric albedo factor files corresponding to the three nearest angles, yielding an interplated matrix $F_0$. The MC\_RSP is then retrieved from the CALDB. On one hand, we apply Equation \ref{Eq1} to obtain $M$, as detailed in subsection \ref{subsect:M}. On the other hand, we interpolate the MC\_RSP files corresponding to the three nearest GRB theta/phi angles, resulting in matrix $F_1$. Finally, after retrieving the energy resolution matrix, the PI channels, the effective high/low thresholds, the total DRM (tDRM) was generated: the sum of $F_1$ and $M$ is then convolved with energy resolution matrix and mapped to PI channels within the effective high and low thresholds.

\begin{figure}[H]
    \centering
   \includegraphics[width=1\linewidth]{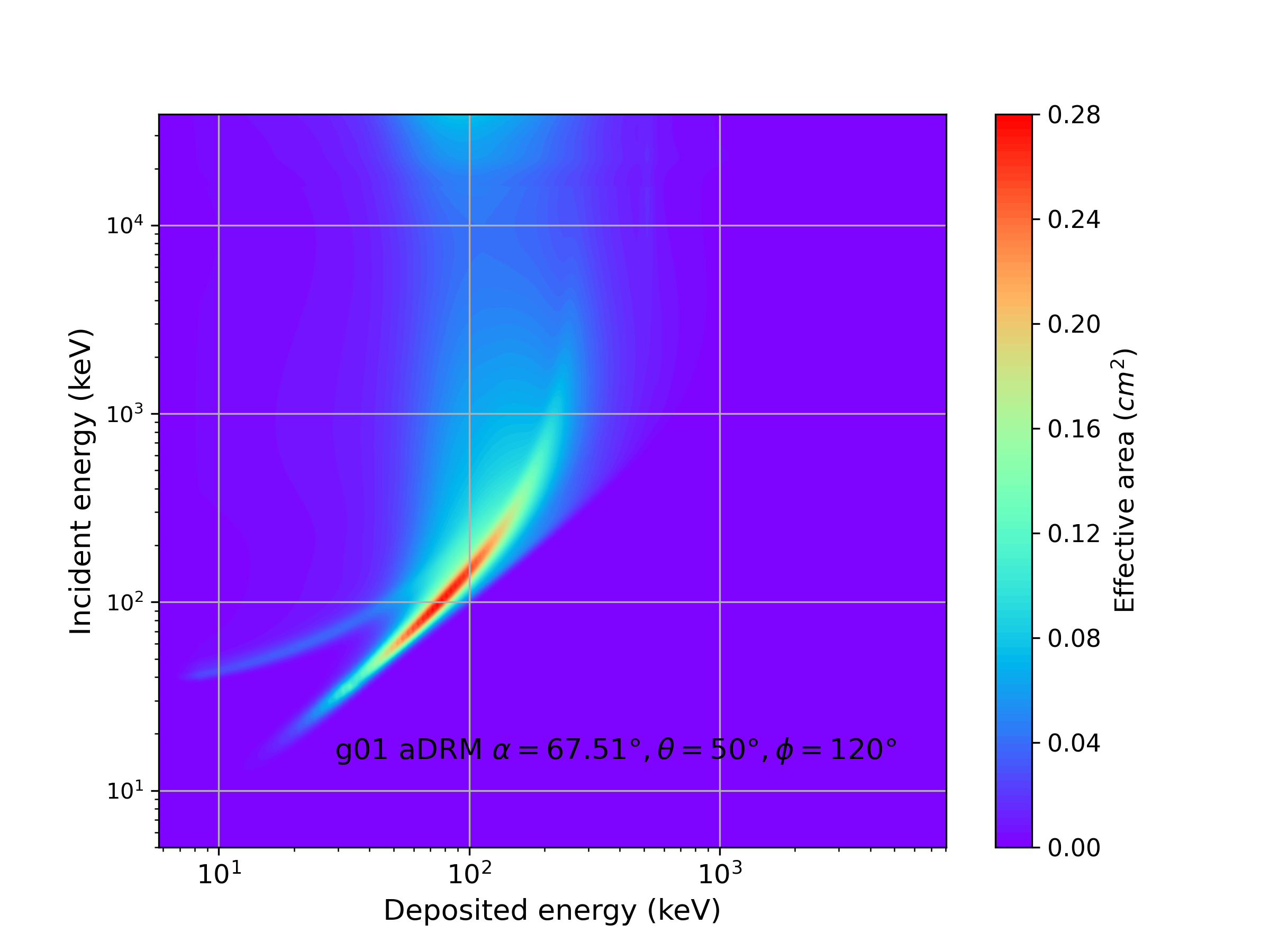}
    \includegraphics[width=1\linewidth]{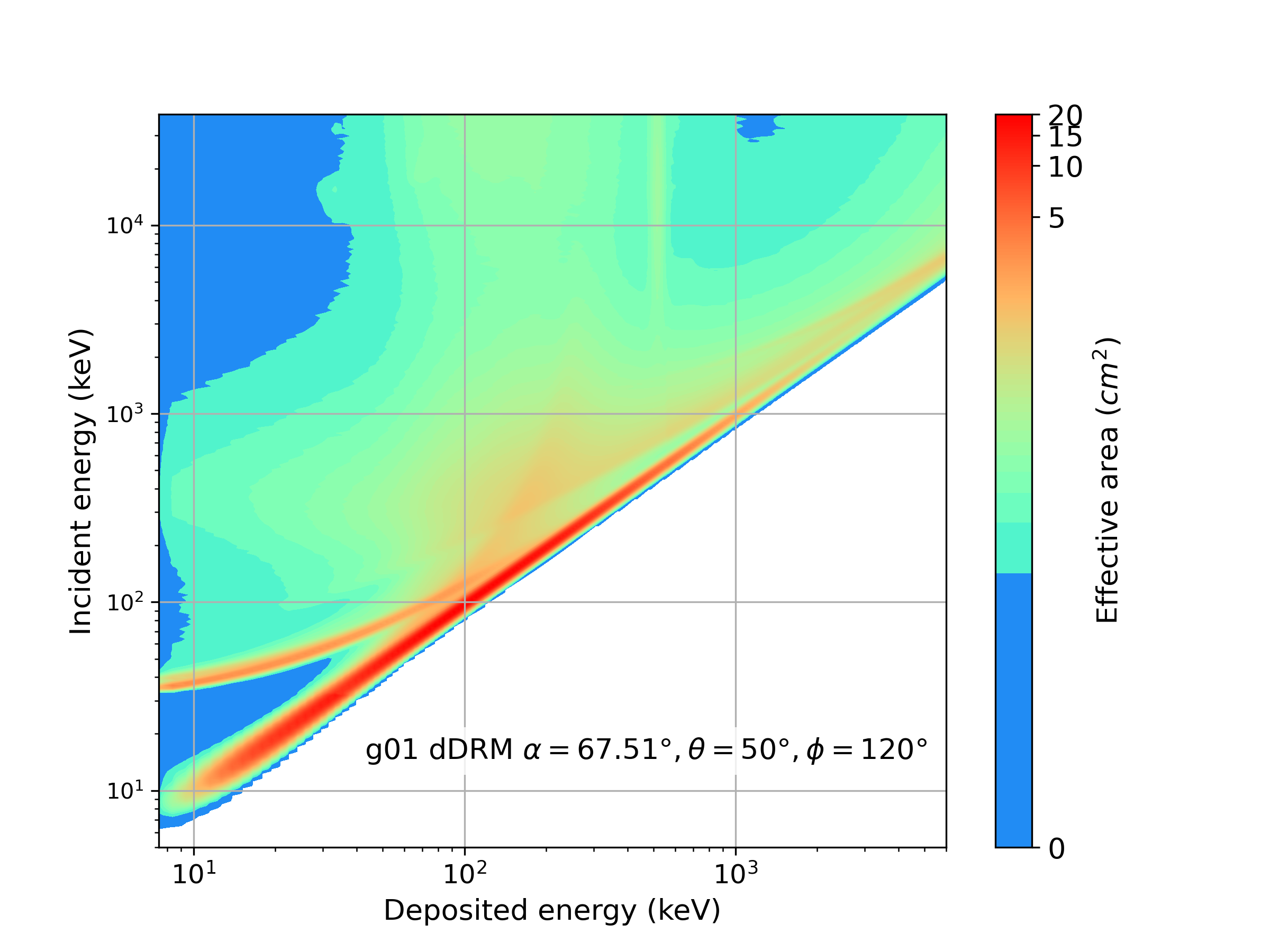}
   \includegraphics[width=1\linewidth]{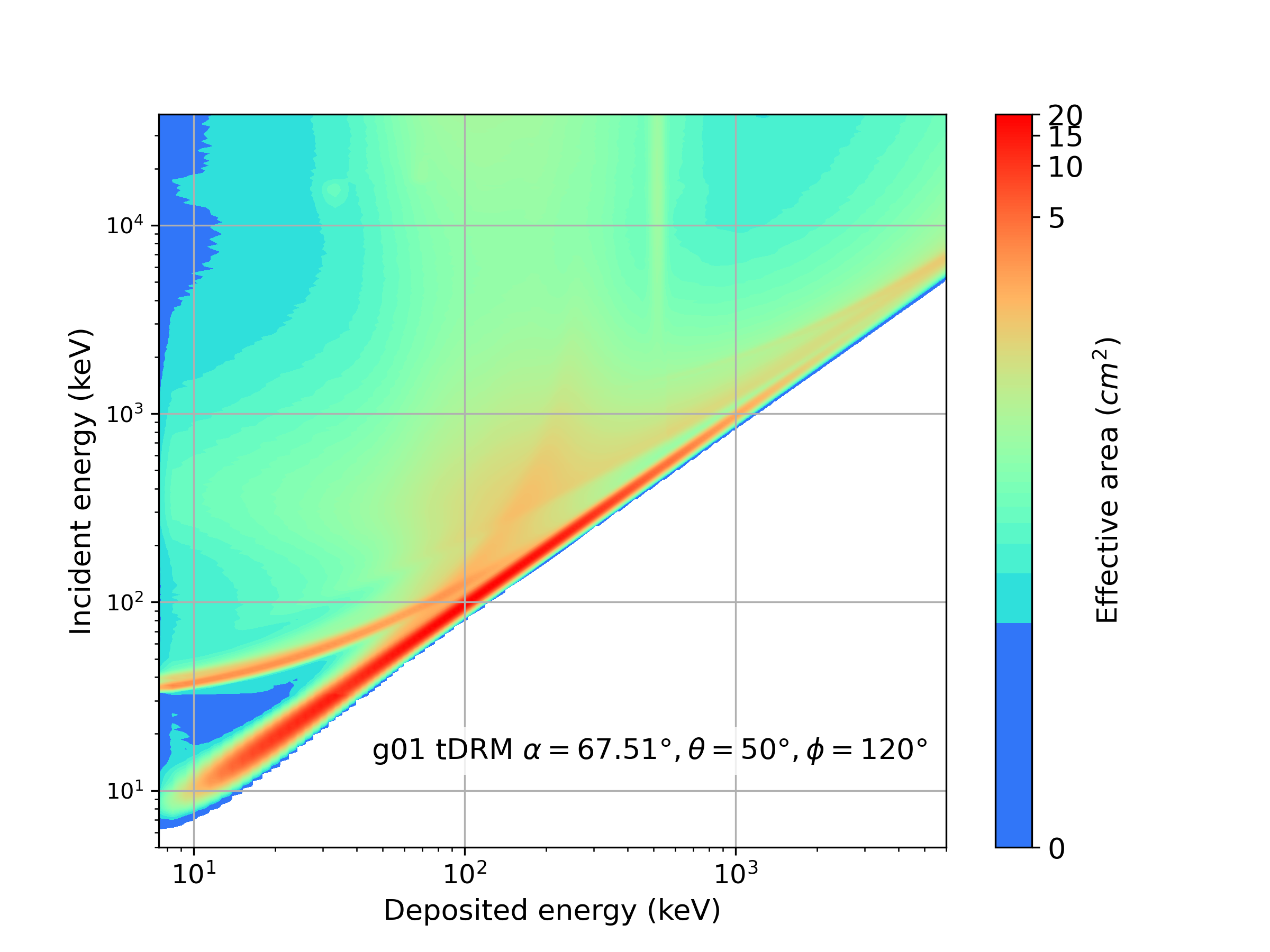}
    \caption{An illustration of albedo DRM (aDRM, top), direct DRM (dDRM, middle) and the total DRM (tDRM, bottom) for GRD01. The incident angle is $\theta=50^{\circ}$ and $\phi=120^{\circ}$. tDRM is the sum of aDRM and dDRM.}
    \label{Fig5a}
\end{figure}

\begin{figure}[H]
    \centering
   \includegraphics[width=1\linewidth]{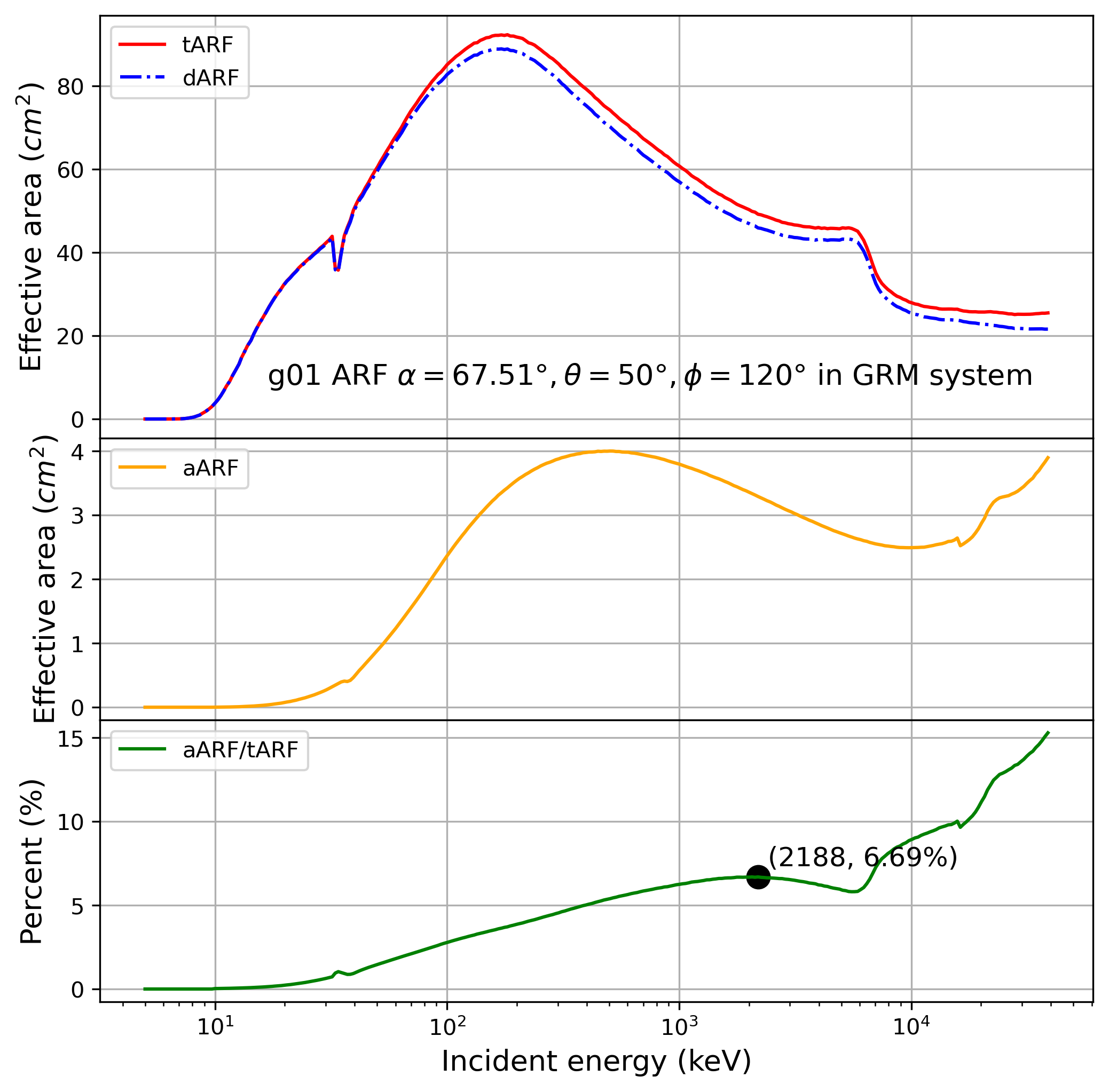}
    \caption{The ARFs of GRD01 (denoted as g01). The top panel shows the dARF (blue dashed line) and the tARF (red line), the middle panel is the aARF, which equals to tARF-dARF (orange line), and bottom panel is the ratio between the aARF and the tARF. The incident angle is $\theta = 50^{\circ}$, $\phi = 120^{\circ}$.}
    \label{Fig5b}
\end{figure}

\begin{figure}[H]
  \centering
   \includegraphics[width=\linewidth]{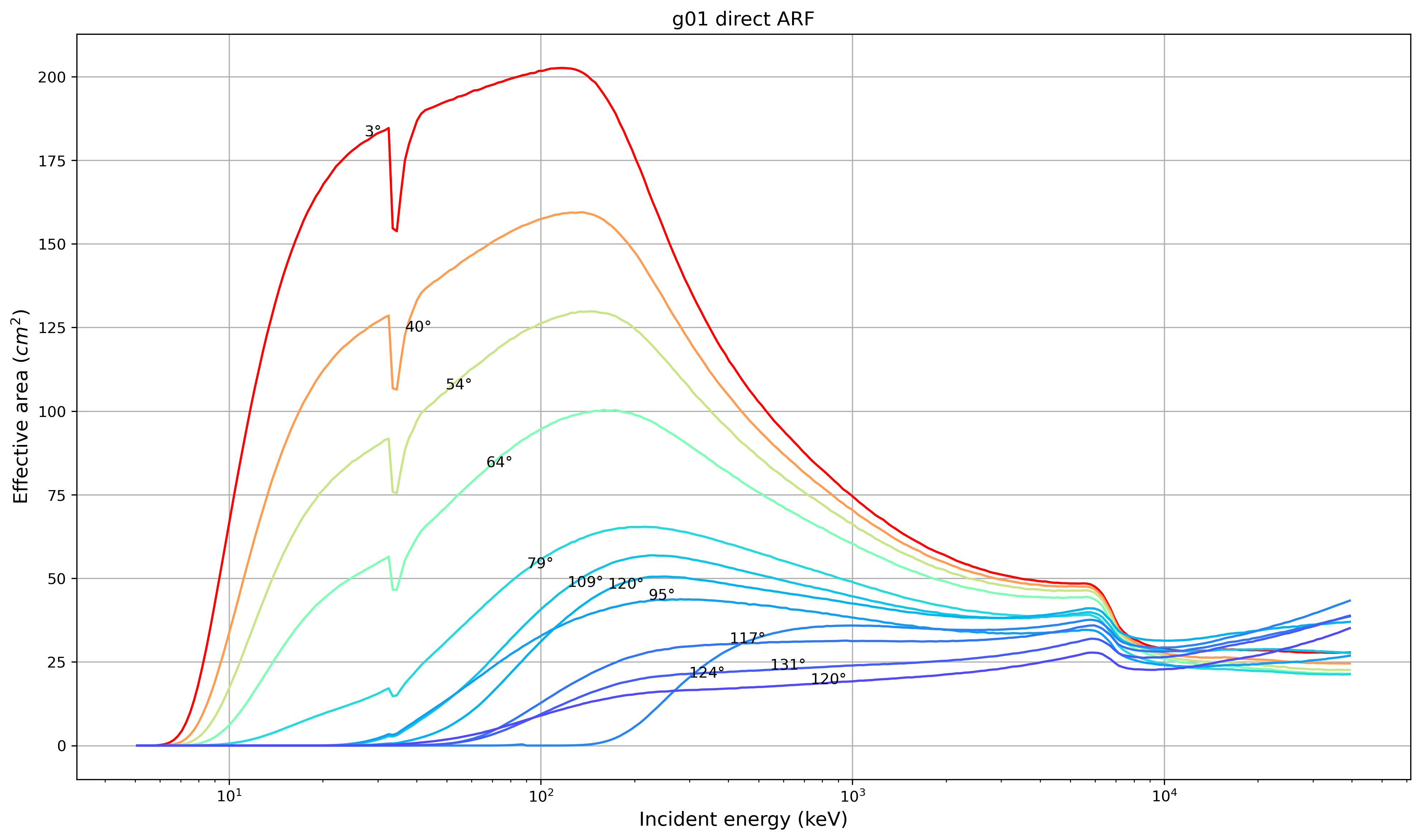}
  	  \caption{The dARF of various incident angles. The angle labeled on the curve represents the angular separation between the GRB and the LoS of GRD01.
  	  }
  	     \label{Fig6}
\end{figure}

\section{Atmospheric albedo effect on energy response}

In this study, we consider a scenario in which the angle $\alpha$ between the GRB and the satellite is $67.51^{\circ}$, the $\theta$ and $\phi$ angles of the GRB in the GRM coordinate system are $50^{\circ}$ and $120^{\circ}$, respectively, which represents a general situation. The aDRM, dDRM and tDRM of this scenario are depicted in Fig.~\ref{Fig5a}, while the albedo, the direct, and total Ancillary Response File (aARF, dARF, and tARF, respectively) are presented in Fig.~\ref{Fig5b}. As shown in Fig.~\ref{Fig5b}, atmospheric scattering primarily influences incident energy greater than tens of keV. The maximum value of aARF/tARF below 5 MeV, along with its corresponding incident energy, is marked in the figure.

During the development of the GRM CALDB, we investigated the effective area of the GRM that is attributable solely to atmospheric scattering. Utilizing 24 hours of STK policy data, which encompasses the SVOM satellite's position and attitude quaternions, we randomly selected 1000 GRB positions, excluding those located in the Earth's shadow. For better clarity of the following context, the dARFs of various incident angles are presented in Fig.~\ref{Fig6}.

Given the anti-solar pointing of the SVOM satellite, Earth occasionally enters the field of view of GRD. Three scenarios are analyzed as follows: 
1) GRD01 is oriented toward the Earth's limb, with an angular separation of 68.03$^\circ$ between the Earth's center and the GRD01 LoS. 
2) GRD01 is directed toward the middle of the sky, with an angular separation of 113.38$^\circ$ between the Earth's center and the GRD01 LoS.
3) GRD01 is anti-Earth-pointing, with an angular separation of 160.69$^\circ$ between the Earth's center and the GRD01 LoS. 

\begin{figure*}
  \centering
   \includegraphics[width=0.48\linewidth]{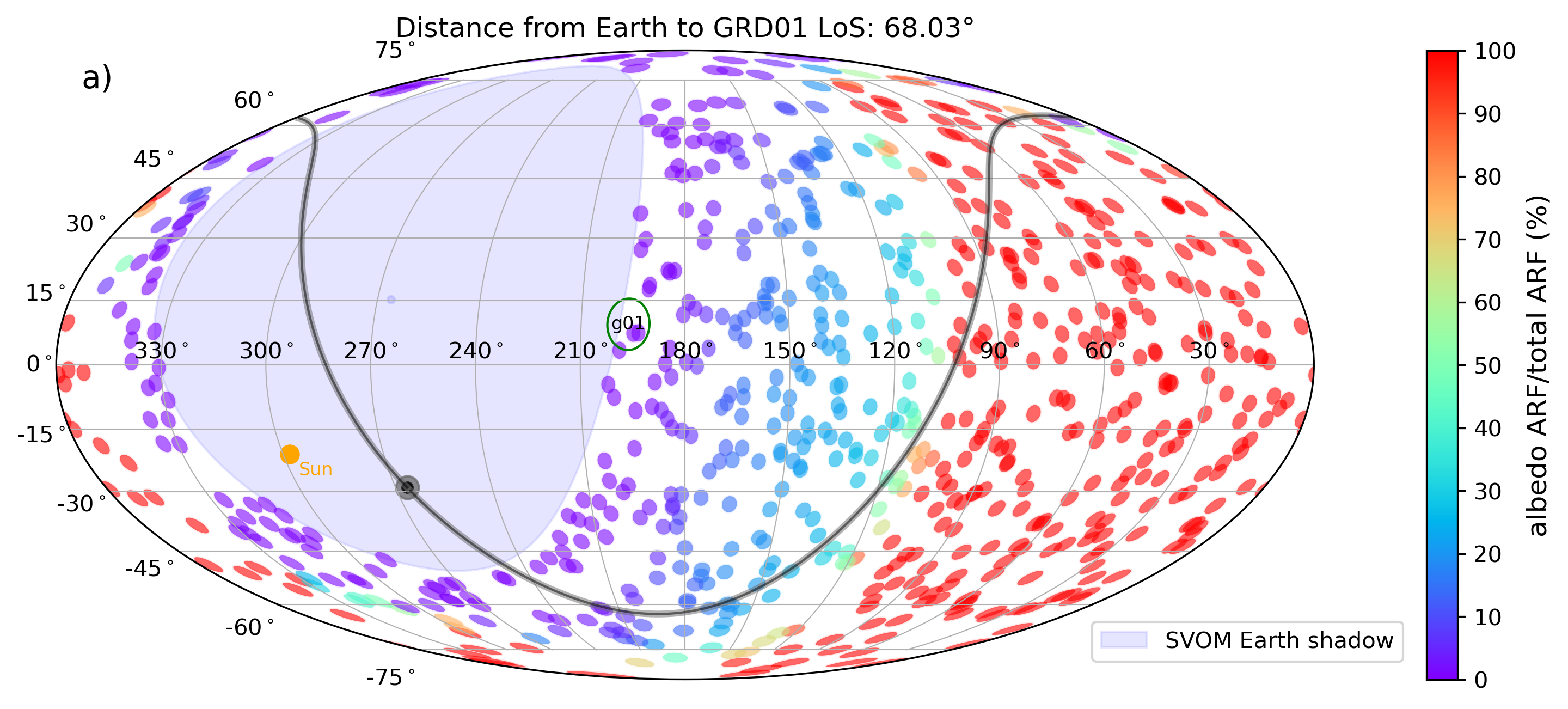}
   \includegraphics[width=0.48\linewidth]{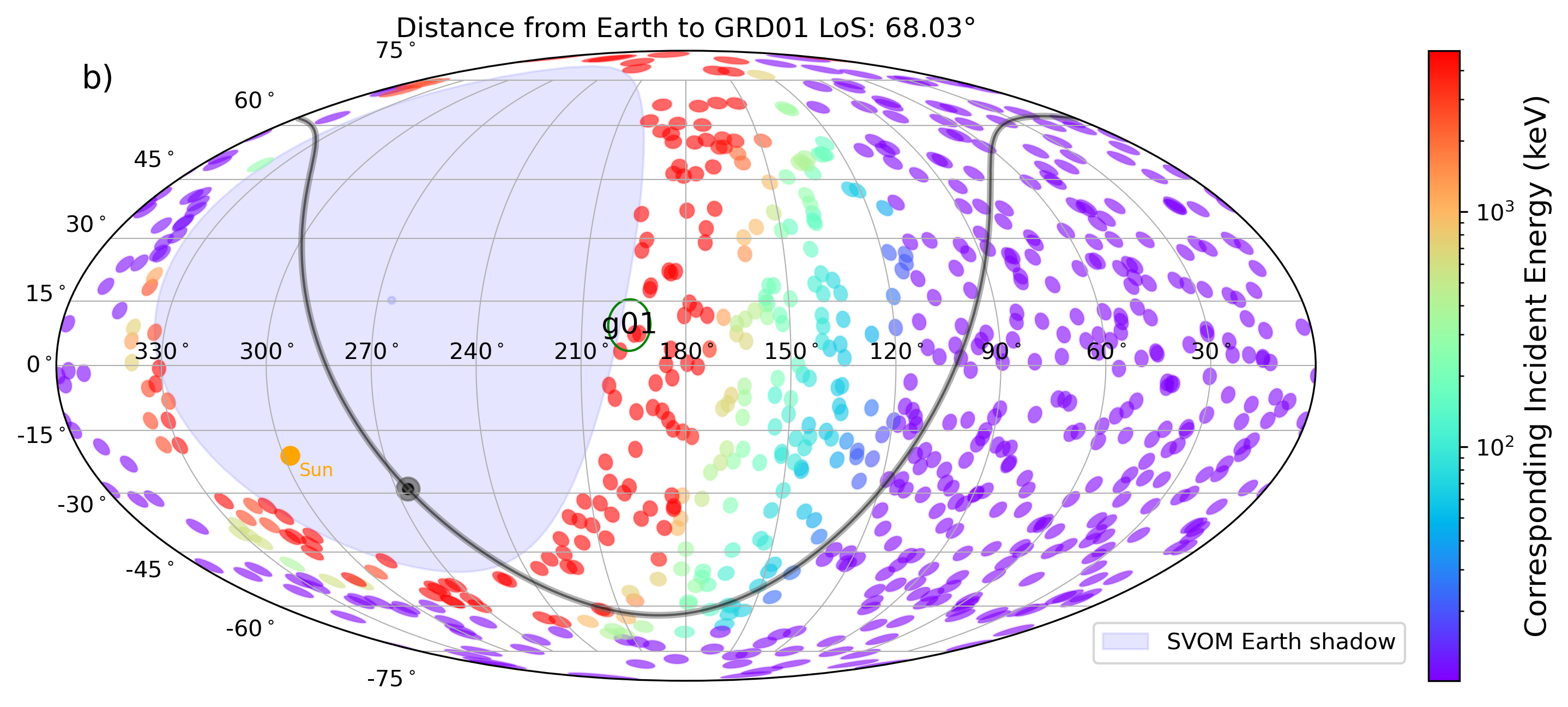}
   \includegraphics[width=0.48\linewidth]{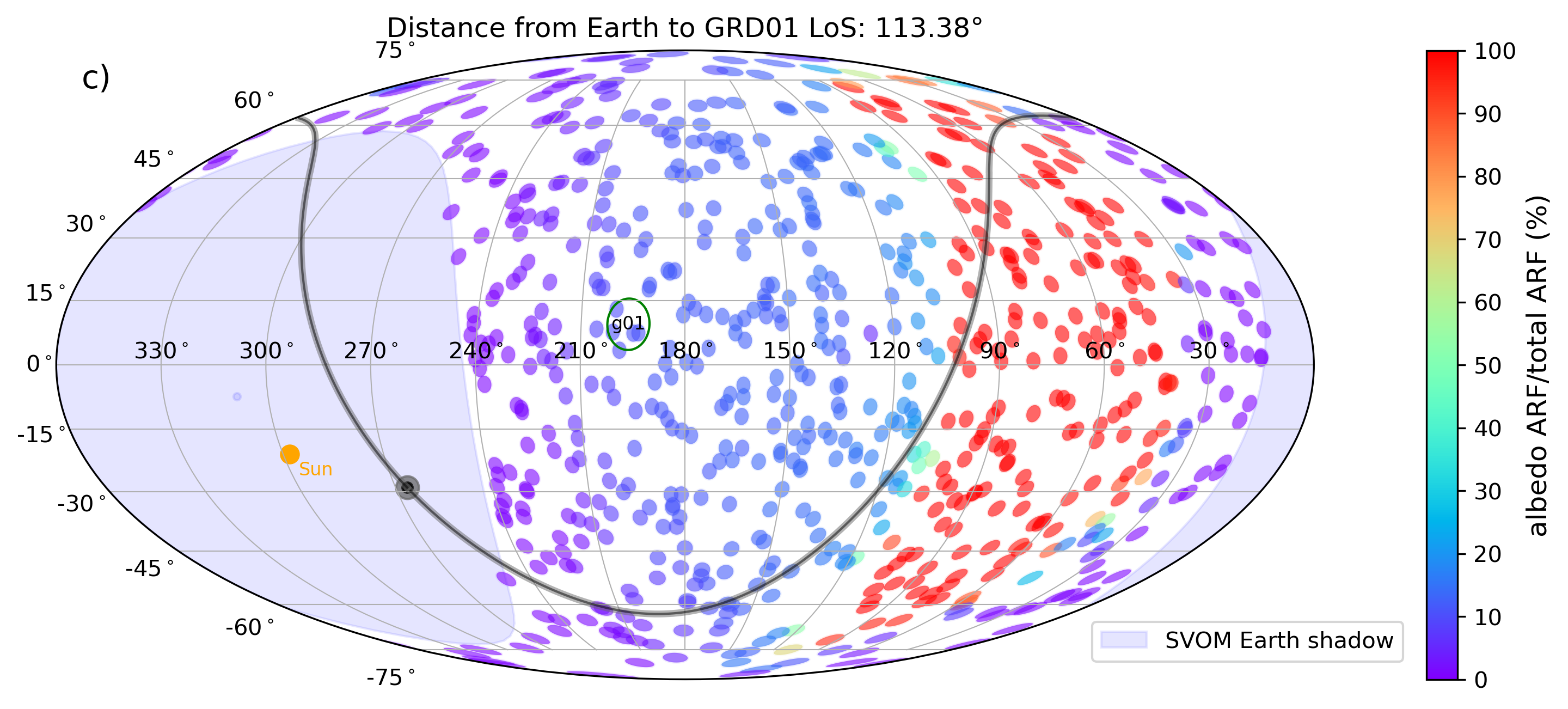}
   \includegraphics[width=0.48\linewidth]{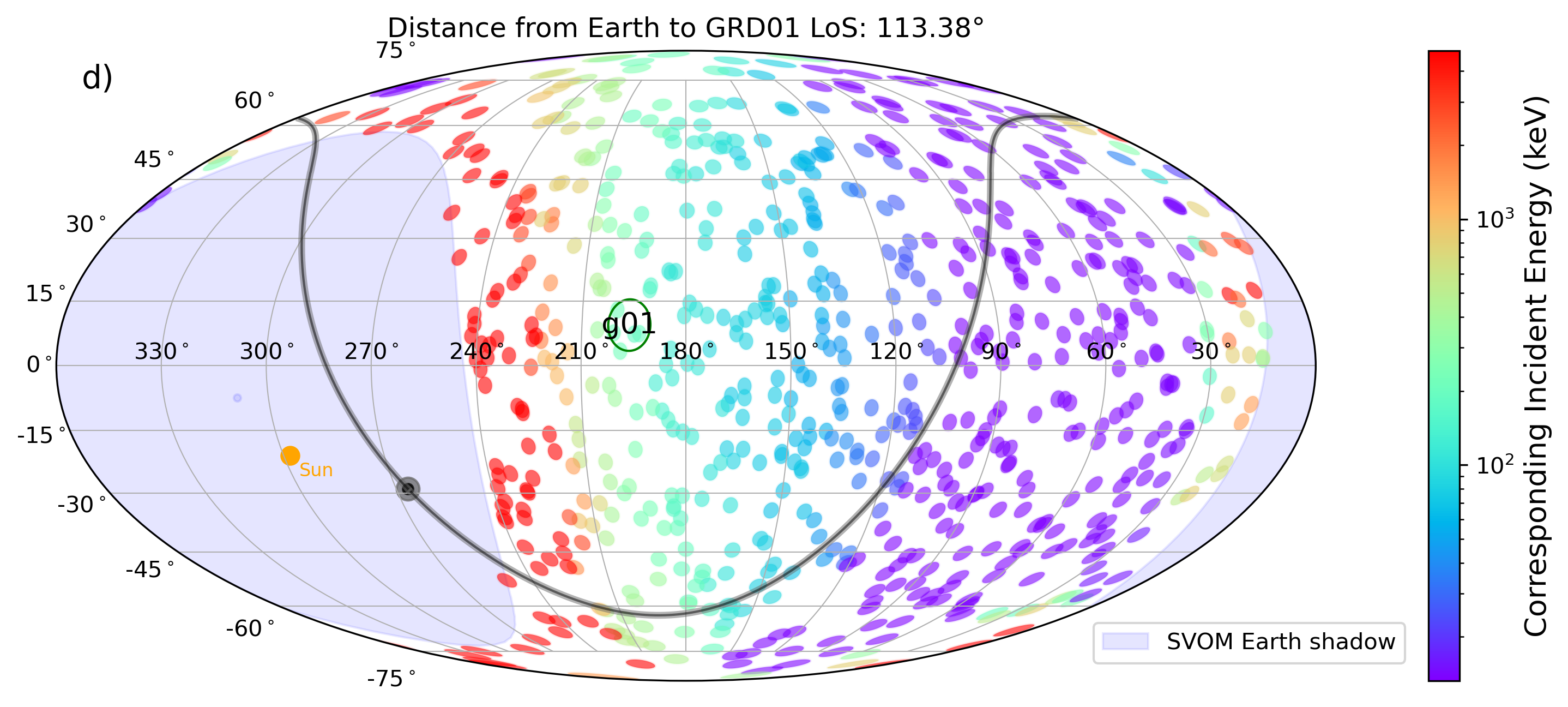}
   \includegraphics[width=0.48\linewidth]{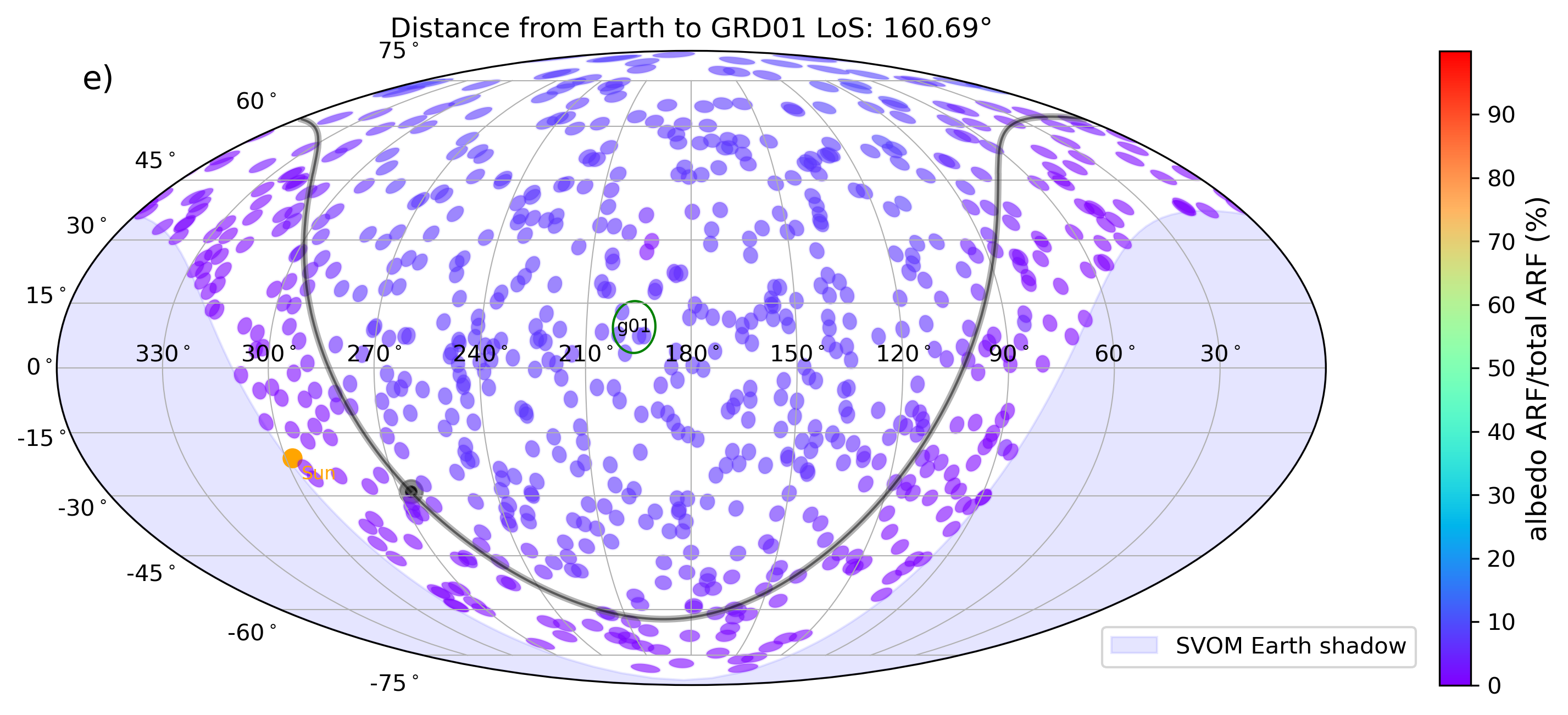}
   \includegraphics[width=0.48\linewidth]{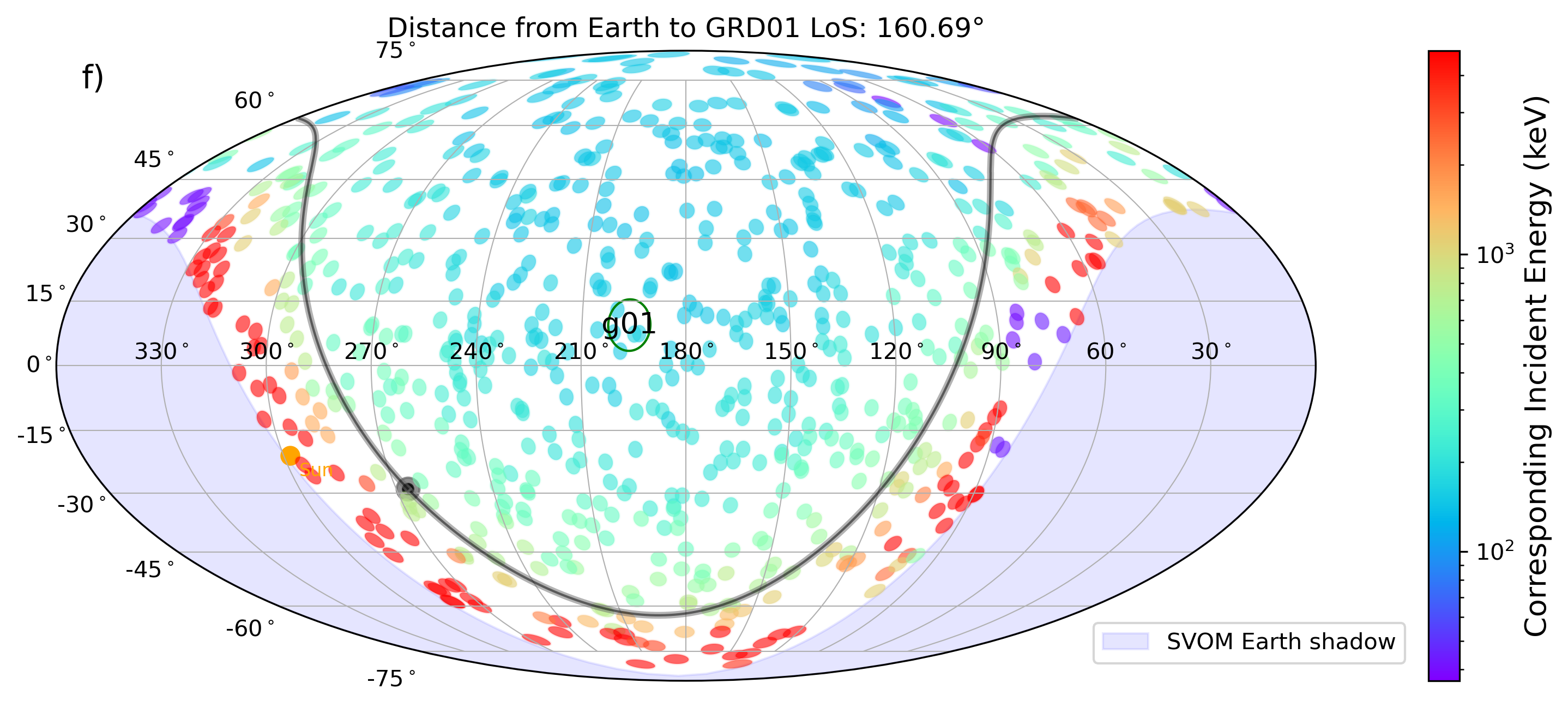}
   \caption{
   Results for three scenarios. The black line represents the Galactic plane, the black dot marks the Galactic center, and the blue shadow indicates the SVOM Earth shadow. The label g01 corresponds to the LoS of the GRD01 probes. The colored circular spots represent randomly sampled GRB locations, with different color codes for each panel: {\bf a) c) e) panel:} the color bar represents the maximum value of the percentage of the aARF relative to the tARF; {\bf b) d) f) panel:} the color bar indicates the corresponding incident energy of the maximum value of aARF/tARF. 
   }
   \label{Fig7}
\end{figure*}

\begin{figure}
  \centering
   \includegraphics[width=1\linewidth]{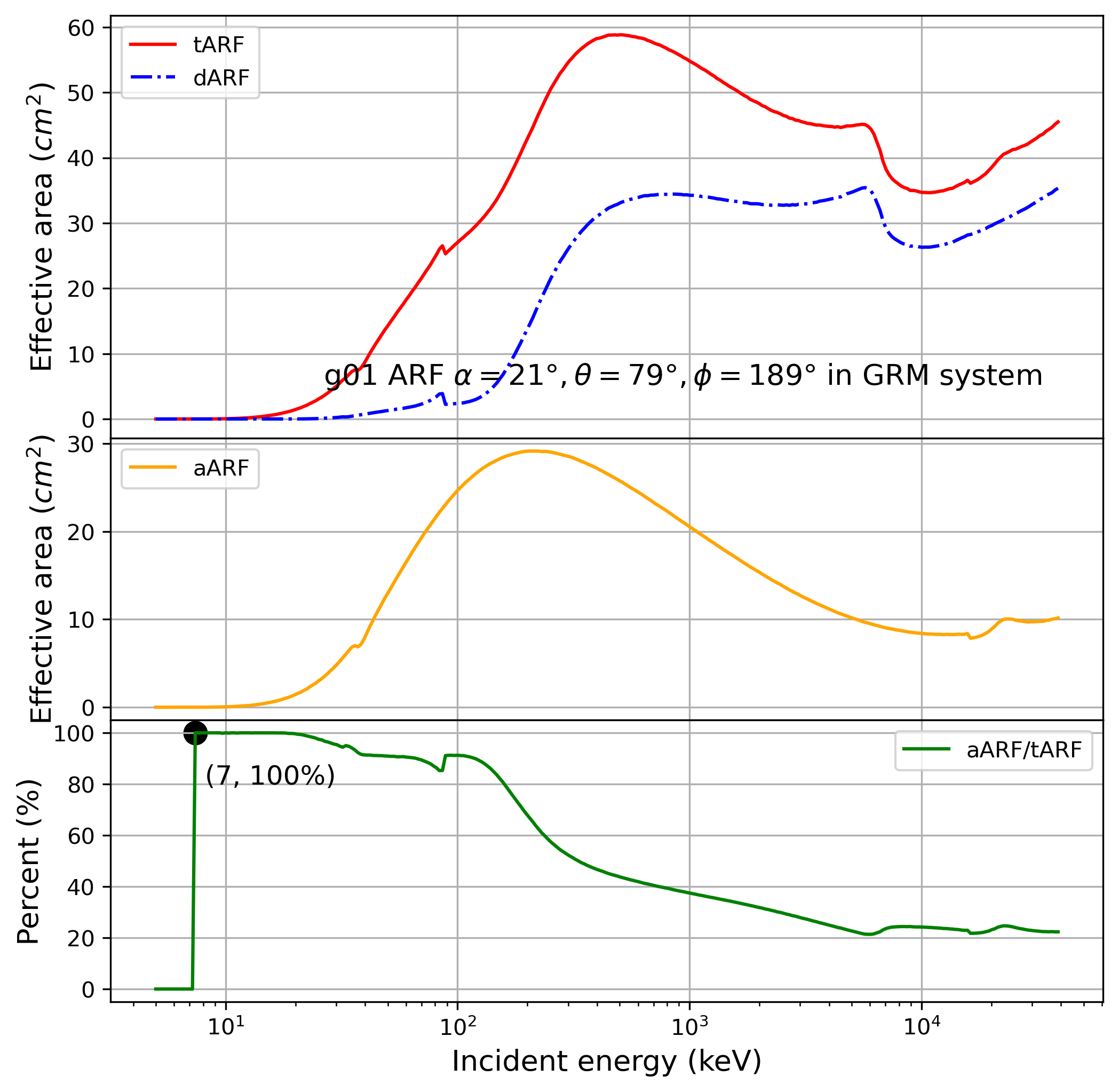}
   \caption{The ARFs of GRM detector g01. The top panel shows the dARF (blue dashed line) and the tARF (red line), the middle panel is the aARF, which equals to tARF-dARF (orange line), and bottom panel is the ratio between the aARF and the tARF. The incident angle is $\theta = 79^{\circ}$, $\phi = 189^{\circ}$.
   }
   \label{Fig7a}
\end{figure}

The results of these scenarios are presented in Fig.~\ref{Fig7}. 
For each case, we calculate the aARF, tARF, and maximum value of the aARF/tARF ratio, along with the corresponding incident energy. These values are then plotted on the sky map. Since the energy range of GRM is 15 keV to 5 MeV, the maximum ratio of aARF/tARF is evaluated within this range. In the worst-case scenario (panel a) b) c) d) of Fig.~\ref{Fig7}), the ratio aARF/tARF can reach 100\%. This occurs when the GRD LoS is not anti-Earth pointing, and the incident angle of the GRB is approximately greater than 90$^\circ$. The corresponding incident energy for the maximum aARF/tARF in this case is typically in the range of tens to hundreds of keV. It should be emphasized here that the majority of the 100\% value falls within the 8 - 20 keV range. the primary reason is the reduced direct effective area (dARF = \SI{0}{\cm\squared}) in this band, caused by a larger GRB injection angle. To better illustrate this 100\% case, we randomly selected a scenario where the maximum value reached 100\% and plotted the corresponding aARF/tARF, as shown in Fig.~\ref{Fig7a}. In contrast, in the best-case scenario (panel e) f) of Fig.~\ref{Fig7}), the maximum ratio of aARF/tARF is significantly lower, approximately 10\%. This occurs when the GRD LoS is anti-Earth pointing and the GRB incident angle is about 90°. In this scenario, the corresponding incident energy for the maximum aARF/tARF is mainly concentrated in the range of hundreds of keV.  

The reason why aARF/tARF behaviors as explained above is that the ratio aARF/tARF is a combined effect of the values of dARF and aARF. When the GRD LoS is anti-Earth pointing, the Earth is located behind or to the rear side of the GRD. GRB photons mainly irradiate the GRD directly, and due to the small incident angle, dARF is relatively larger, as shown in Fig.~\ref{Fig6}. In contrast, the probability of atmospheric albedo photons scattering onto the GRD is relatively low, and dARF plays a dominant role in tARF. When the GRB incident angle is approximately 90 degrees, this ratio reaches its maximum in this geometric configuration, about 10\%. When the GRD LoS is not anti-Earth pointing and the GRB incident angle exceeds 90 degrees, dARF becomes relatively smaller, as shown in Fig.~\ref{Fig6}, while the probability of atmospheric albedo photons scattering onto the GRD increases. In this geometric configuration, aARF and dARF are comparable because the dARF at large incident angles is smaller (its maximum is less than about \SI{40}{\cm\squared}). As the incident angle varies, aARF (with a maximum of up to \SI{30}{\cm\squared}) can even dominate tARF, causing the ratio to reach a maximum of up to 100\% in this geometric configuration.

\section{Conclusions}
\label{sect:conclusion}

The atmospheric albedo effect plays a critical role in GRB astronomy, particularly in spectral and localization analyzes. Within the SVOM GRM CALDB, comprehensive simulations and calculations demonstrate that the maximum albedo effective area can reach 100\% of the total effective area when the GRD LoS is not anti-Earth oriented. This is especially pronounced in the 8 - 20 keV range, where the direct effective area drops to zero due to the large GRB injection angle. Under such conditions, the albedo effect must be incorporated into the analysis. In contrast, when the LoS of the GRD is nearly anti-Earth pointing and the GRB incident angle nearly perpendicular to the GRD LoS, the maximum albedo effective area accounts for nearly 10\% of the total effective area.

\begin{acknowledgements}
The Space-based multi-band Variable Objects Monitor
 (SVOM) is a joint Chinese-French mission led by the Chinese
 National Space Administration (CNSA), the French Space
 Agency (CNES), and the Chinese Academy of Sciences
 (CAS). We gratefully acknowledge the unwavering support of
 NSSC, IAMCAS, XIOPM, NAOC, IHEP, CNES, CEA,
 and CNRS. 
The authors are grateful for support from the National Key R$\&$D Program of China (grant Nos. 2024YFA1611701, 2024YFA1611700). 
This work is supported by 
the National Natural Science Foundation of China (Grant No. 
12494572,
12273042
), 
the Strategic Priority Research Program of the Chinese Academy of Sciences (Grant No.
XDB0550300
). The published material is based on observations performed by the SVOM mission funded by the respective space agencies (CNSA and CNES).
\end{acknowledgements}

\bibliographystyle{raa}
\bibliography{bibtex}

\label{lastpage}
\end{document}